\documentclass[a4paper,fleqn,usenatbib]{mnras}
\usepackage{ae,aecompl}
\usepackage{graphicx}	
\usepackage{amsmath}	
\usepackage{amssymb}	

\usepackage{ulem}

\def\simlt{\mathrel{\hbox{\rlap{\hbox{\lower4pt\hbox{$\sim$}}}\hbox{$<$}}}}
\def\simgt{\mathrel{\hbox{\rlap{\hbox{\lower4pt\hbox{$\sim$}}}\hbox{$>$}}}}

\newcommand{\myapprox}{\mathord{\approx}}

\title[Screening from detailed balance]{Screening of fusion reactions from the principle of detailed balance and application to the $pep$ reaction}

\author[D. Kushnir, E. Waxman and A. I. Chugunov]{
Doron Kushnir$^{1}$\thanks{E-mail: doron.kushnir@weizmann.ac.il},
Eli Waxman$^{1}$ and Andrey I. Chugunov$^{2}$
\\
$^{1}$Dept. of Particle Phys. \& Astrophys., Weizmann Institute of
Science, Rehovot 76100, Israel\\
$^{2}$Ioffe  Institute, Polytekhnicheskaya 26, 194021
St.-Petersburg, Russia
}

\date{Accepted XXX. Received YYY; in original form ZZZ}

\pubyear{2018}

\begin{document}
\label{firstpage}
\pagerange{\pageref{firstpage}--\pageref{lastpage}}
\maketitle

\begin{abstract}
Dewitt et al. (1973) derived a useful relation between the plasma screening factor for a reaction of two fusing ions and their chemical potentials, based on the plasma pair distribution functions. We show that their result can be derived in a simpler, more straightforward way, by applying the principle of detailed balance, which also enables us to generalize the relation to reactions involving $N$ fusing ions. In order to demonstrate the usefulness of applying the principle of detailed balance, we calculate the screening factor for the $pep$ reaction, $p+e+p\rightarrow ^{2}$D $+\nu_{e}$. For the plasma conditions near the centre of the Sun, the reaction is suppressed by roughly the same amount by which the reaction $p+p\rightarrow ^{2}$D $+$ $e^{+}+\nu_{e}$ is enhanced. This effect may be measured in the near future.
\end{abstract}




\section{Introduction}
\label{sec:Introduction}

Non-ideal gas corrections to nuclear reaction rates, known as screening corrections, are significant for sufficiently dense plasmas \citep[see, e.g.,][and references therein]{Itoh_etal90,Yakovlev2006,Calder2007,Chugunov2007,Clerouin2018}. The state of a neutral plasma composed of ions with mass $m$ and charge $Ze$, which are embedded in a background of electrons, can be described by its temperature, $T$, and ion number density, $n$. Alternatively, the two dimensionless parameters $\Gamma=\beta(Ze)^2/a$ and $\tau=[27\pi^4\beta m(Ze)^4/h^2]^{1/3}$ can be used \citep[e.g.,][]{Salpeter1969}. Here, $\beta=1/T$, $a=(3/4\pi\rho)^{1/3}$ is the mean ionic distance, and $h$ is Plank's constant. The parameter $\Gamma$ measures the strength of the coupling of the plasma, and $\Gamma/\tau$ measures the importance of quantum corrections. Similar parameters can be introduced for multi-ion plasmas \citep[e.g.,][]{Itoh_etal90}. The screening corrections for the $\Gamma/\tau\ll1$ case, which is the most relevant for thermonuclear burning in ordinary stars and in supernovae, was first calculated by \citet{Salpeter1954}, and raised some controversy over the years \citep[see, for example][]{Bahcall2002}. \citet{Dewitt1973} derived a relation between the plasma screening factor for a reaction of two fusing ions and their chemical potentials. Using the pair distribution functions of the plasma, under the assumption that the ions can be described by classical statistics, they have obtained
\begin{eqnarray}\label{eq:screening factor dewitt}
f=\exp\left[\beta\left(\mu^{C}(Z_{1})+\mu^{C}(Z_{2})-\mu^{C}(Z_{1}+Z_{2})\right)\right],
\end{eqnarray}
where $Z_{1}$, $Z_{2}$ are the ion charges, $\mu^{C}$ is the Coulomb interaction contribution to the chemical potential ($\mu_i^{C}=\partial_{N_i} F^C$, and $F^C$ is the Coulomb interaction contribution to the Helmholtz free energy). Later on, \citet{{Yakovlev1989}} provided a somewhat different derivation of Equation~\eqref{eq:screening factor dewitt}, and also provided the screening factor for the triple-$\alpha$ reaction, with $\beta\left(3\mu^{C}(Z=2)-\mu^{C}(Z=6)\right)$ as the argument of the exponent. A physical interpretation of Equation~\eqref{eq:screening factor dewitt} was given by \citet{Gruzinov1998}. In brief, $\mu^{C}(Z_{1})+\mu^{C}(Z_{2})-\mu^{C}(Z_{1}+Z_{2})$ is the Coulomb work, $\delta W$, done by the plasma as the ions $Z_{1}$, $Z_{2}$ are removed and the composite ion $Z_{1}+Z_{2}$ is added. Thus, the relative kinetic energy of the fusing ions is lower by $\delta W$ than in the absence of screening. Therefore, the probability of fusing is increased by a factor $\exp(\beta\delta W)$.

Equation~\eqref{eq:screening factor dewitt} is important for two reasons. First, it should remove any controversy, if still remains, regarding Salpeter's formula \citep{Bruggen1997}. Secondly, it is very useful as it allows one to calculate the screening factor based on the Helmholtz free energy, which is often known to a high level of accuracy \citep[see, e.g.,][]{Potekhin2000,Potekhin2009,Chugunov2009b}.

In this paper, we provide a simple derivation of Equation~\eqref{eq:screening factor dewitt} by applying the principle of detailed balance (Section~\ref{sec:proof}). Our derivation allows us to generalize Equation~\eqref{eq:screening factor dewitt} to $N$ fusing ions \citep[the generalization can also be done by following the arguments of][]{Gruzinov1998}. We also briefly comment on plasma conditions under which $\Gamma/\tau\sim1$, where the screening factor of Equation~\eqref{eq:screening factor dewitt} is multiplied by a correction factor, known as a 'quantum' correction.

To demonstrate the usefulness of applying the principle of detailed balance, we calculate in Section~\ref{sec:pep} the screening factor of the $pep$ reaction, $p+e+p\rightarrow ^{2}$D$+\nu_{e}$ (we are unaware of earlier derivations). For the conditions near the centre of the Sun, the $pep$ reaction is suppressed by a few percent. The $pep$ solar neutrino flux has been measured to $\sim20\%$ accuracy \citep{Bellini2012,Agostini2017}, and the required accuracy to measure the $pep$ screening suppression may be reached in the near future \citep{Newstead2018}.

\section{Deriving the screening correction from the principle of detailed balance}
\label{sec:proof}

Consider a reaction involving $N$ ion reactants with charges $Z_{i}$ ($i=1,..,N$) and masses $m_{i}$. We are interested in determining the screening correction to the reaction rate due to the presence of a plasma, under the assumptions that (i) Coulomb interactions make a small contribution to the free energy of the electrons, (ii) all plasma ions are non-degenerate, and (iii) $\Gamma/\tau\ll1$, meaning that the length scales relevant for the nuclear process (specifically the de Broglie wavelength of the fusing ions and the tunnelling length of particles with Gamow-peak energy) are much smaller than the Debye length, over which plasma screening takes place. Under these conditions, the screening correction (for given plasma parameters) would be the same as for some hypothetical $N$ ions with similar charges, $Z_{i}$, masses, $m_{i}$, and velocity distributions, that interact to produce a single ion of charge $Z_j=\sum_iZ_i$ and a photon, $\sum_i n_i\rightarrow n_j+\gamma$, and that the density of which is negligible. The plasma properties are not affected by the presence of such ions with negligible density, and the screening correction to this reaction would be identical to the screening correction of the reaction in which we are interested.

Two comments are in place here. The first is related to the assumption that the nuclear interaction length scale is much shorter than the plasma screening length scale. In order to derive the screening correction, we will assume below that the inverse hypothetical reaction, $n_j+\gamma\rightarrow\sum_i n_i$, is not affected by the plasma. This is valid as long as the screening of the ions in the final state is not important, which holds for the case where the nuclear scale is much shorter than the Debye scale. Note also that in this case 'quantum' corrections are negligible, i.e. tunnelling is unaffected by screening. The inclusion of 'quantum' corrections is briefly discussed at the end of this section.

The second comment is related to the motivation for using 'test ions' of negligible density. In order to derive the screening factor, we will consider a situation in which the interacting ions are in thermal equilibrium. The relative abundances of the ions may differ in thermal equilibrium from the ion abundances in the plasma under consideration, which, in turn, may modify the plasma screening. Using 'test ions' of negligible density implies that the properties of the plasma are not affected by their relative abundances. For non-degenerate ions, the energy and velocity distributions of the 'test ions' would be similar to those of the plasma ions in which we are interested.

Consider, therefore, a thermal equilibrium of the 'test ions', in which the reaction $\sum_i n_i\leftrightarrow n_j+\gamma$ is in balance, such that
\begin{equation}\label{eq:detailed balance}
\prod_{i=1}^{N}n_{i}\langle\sigma v\rangle_{f}=n_{j}n_\gamma\langle\sigma v\rangle_{r}.
\end{equation}
Here, $\langle\sigma v\rangle_{f}$ is the forward reaction cross-section (including factors for identical ions), and $\langle\sigma v\rangle_{r}$ is the reverse reaction cross section (note that $n_{\gamma}$ is usually included in $\langle\sigma v\rangle_{r}$, but for clarity we keep the $n_{\gamma}$ term).
Using Equation~(\ref{eq:detailed balance}), we may write
\begin{equation}\label{eq:sigma_f}
  \langle\sigma v\rangle_{f}=\frac{n_jn_\gamma}{\prod_{i=1}^{N}n_{i}}\langle\sigma v\rangle_{r}.
\end{equation}
This enables us to derive the screening correction by expressing the densities $n_k$ in terms of the chemical potentials $\mu_k$, which satisfy
\begin{equation}\label{eq:chemical potentials}
\sum_{i=1}^{N}\mu_{i}=\mu_{j}.
\end{equation}
Writing the free energy as a sum of the free energy of an ideal gas and a small Coulomb correction, $F=F^{\rm id}+F^C$, we have $\mu_i=\partial_{N_i} F = \mu_i^{\rm id}+\mu_i^C$. For non-degenerate ions,
\begin{equation}\label{eq:mu kin}
\mu_{i}^{\rm id}=m_i c^2 + T \ln\left(\frac{n_{i}}{n_{i0}}\right),
\end{equation}
where $n_{i0}$ depends on $T$ and $m_i$. For example, in the non-relativistic case
\begin{equation}\label{eq:ni0}
n_{i0}\equiv g_{i}\left(\frac{2\pi m_{i}T}{ h^{2}}\right)^{3/2},
\end{equation}
where $g_{i}$ is the effective number of internal degrees of freedom of ion $i$. Using eqs.~(\ref{eq:sigma_f}), (\ref{eq:chemical potentials}) and (\ref{eq:mu kin}) we may write
\begin{equation}\label{eq:detailed balance 2}
\langle\sigma v\rangle_{f}=\left[\frac{n_{j0}n_\gamma}{\prod_{i=1}^{N}n_{i0}}
e^{\beta q}\langle\sigma v\rangle_{r}\right]
\exp\left[\beta\left(\sum_{i=1}^{N}\mu_{i}^{C}-\mu_{j}^{C}\right)\right],
\end{equation}
where $q=\sum_{i=1}^{N}m_{i}c^{2}-m_{j}c^{2}$. The first term in square brackets depends only on the plasma temperature and is independent of its density (the reverse reaction is not screened, hence $\langle\sigma v\rangle_{r}$ depends only on $T$; here we are using assumption (iii) as explained above). The second, exponential, term depends on the density and approaches unity as the density vanishes.
 We may, therefore, identify the screening correction as
\begin{equation}\label{eq:screening factor}
f=\exp\left[\beta\left(\sum_{i=1}^{N}\mu_{i}^{C}-\mu_{j}^{C}\right)\right].
\end{equation}
Finally, we note that $\mu_{i}^{C}$ is the same for the test ions and for the interacting ions in which we are interested, since for non-degenerate ions the Coulomb contribution to the free energy depends only on the charge.

The quantum corrections are associated with the screening modification of the shape of the interionic potential on a length scale comparable to the tunnelling length scale, which affects the tunnelling probability \citep[see, e.g.,][]{Itoh_etal90,Chugunov2007,Chugunov2009}. To satisfy detailed balance for reactions that take place via a compound nucleus, quantum corrections should be included not only for the entrance channel, but also for all the decay channels of the respective state of the compound nucleus. This leads to a modification of the branching ratios of the decay channels in the presence of a plasma. Denoting the partial width of the $n$th decay channel in vacuum by $\gamma_n$ (with $n=0$ for the entrance channel), and the total width by $\Gamma=\sum_n \gamma_n$, the quantum tunnelling correction for the $n$th channel, $\mathfrak f_n^\mathrm{q}$ (typically $\mathfrak f_n^\mathrm{q}<1$), modifies the tunnelling probability and hence the partial width, which becomes $\tilde \gamma_n=\mathfrak f_n^\mathrm{q}\gamma_n$. The total width becomes $\tilde \Gamma=\sum \tilde \gamma_n$, leading to a branching ratio for the $n$th channel $\tilde b_n=\tilde \gamma_n/\tilde\Gamma$, which differs from the value in vacuum, $b_n=\gamma_n/\Gamma$. As a result, the reaction rate for a given entrance and exit channel is modified by a factor
\begin{equation}
 f^\mathrm{q}_{0,n}=\mathfrak f_0^\mathrm{q} \frac{\tilde b_n}{b_n}=\mathfrak f_0^\mathrm{q} \mathfrak f_n^\mathrm{q} \frac{\Gamma }{\tilde \Gamma},\label{quanCorr}
\end{equation}
which should be applied in addition to the factor $f$ given by Eq.~(\ref{eq:screening factor}).%
\footnote{For reactions with more than one charged particle at the exit channel, $\mu_j^C$ in equation \eqref{eq:screening factor} corresponds to the Coulomb part of the compound nucleus chemical potential.}
 That is, the reaction rate in the presence of a plasma should be multiplied by the combined (classical+quantum) screening factor $f  f^\mathrm{q}_{0,n}$. Here, the factor $\mathfrak f_0^\mathrm{q}$ corrects the probability of the compound nucleus formation and $\tilde b_n/b_n$ originates from the modification by quantum corrections of the probability of decay to a given channel $n$.
The quantum correction factor in this form obviously agrees with the principle of detailed balance, as it is just the same for the direct and for the reverse reactions. We plan to consider the calculation of the $\mathfrak f^\mathrm{q}_n$ factors in a subsequent publication.

\section{The screening correction for the $pep$ reaction}
\label{sec:pep}

We would like to calculate the screening factor for the $pep$ reaction, $p+e+p\rightarrow ^{2}$D$+\nu_{e}$ \citep{Schatzman1958,Bahcall1969}. We first derive the correction for the case of non-degenerate electrons with de Broglie wavelengths much smaller than the Debye length. Let us consider then test particles $\tilde{p}$, $\tilde{e}$, $\tilde{^{2}\textrm{D}}$ and $\tilde{\nu}_{e}$ with the relevant properties and negligible density \footnote{Alternatively, one may imagine particle physics in which lepton number is not a conserved quantum number, such that the reaction  $\tilde{p}+\tilde{e}+\tilde{p}\rightarrow \tilde{^{2}\textrm{D}}+\gamma$ is possible, and repeats the exact same derivation from Section~\ref{sec:proof}.}. In thermal equilibrium,
\begin{equation}\label{eq:detailed balance pep}
n_{\tilde{p}}^{2}n_{\tilde{e}}K_{pep}(n_e,n_p,T)=n_{\tilde{\rm{D}}}n_{\tilde{\nu}}\langle\sigma v\rangle_{r},
\end{equation}
which implies for the rate parameter
\begin{equation}\label{eq:detailed balance pep2}
K_{pep}=\frac{n_{\tilde{\rm{D}}}n_{\tilde{\nu}}}{n^2_{\tilde{p}}n_{\tilde{e}}}\langle\sigma v\rangle_{r}.
\end{equation}
Following the same derivation of Section~\ref{sec:proof}, we find
\begin{equation}\label{eq:detailed balance 3}
K_{pep}=\left[\frac{ n_{D0} n_{\nu0}}{n_{e0}n_{p0}^2}
e^{\beta q}\langle\sigma v\rangle_{r}\right]
\exp\left[\beta\left(2\mu_p^C+\mu_e^C-\mu_D^C\right)\right],
\end{equation}
where $q=(2m_p+m_e-m_D)c^2$ and
\begin{equation}\label{eq:mu nu}
n_{\nu0}=16\pi\left(\frac{T}{ch}\right)^{3}.
\end{equation}
The first term in square brackets depends only on the plasma temperature and is independent of its density (the reverse reaction is not screened, hence $\langle\sigma v\rangle_{r}$ depends only on $T$). The second, exponential, term depends on the density and approaches unity as the density vanishes. Noting that $\mu_p^C=\mu_D^C$, we may identify the screening correction as
\begin{eqnarray}\label{eq:pep screening factor}
f_{pep}=\exp\left[\beta\left(\mu_{p}^{C}+\mu_{e}^{C}\right)\right].
\end{eqnarray}

This derivation is accurate for the case of non-degenerate electrons with de Broglie wavelengths much smaller than the the Debye length. For the plasma conditions near the centre of the Sun, neglecting degeneracy is a good approximation but the de Broglie wavelength of the electrons is comparable to Debye's length, $\lambda_{D}\sim h/\sqrt{2\pi m_{e}T}$, i.e. the assumption that $\lambda_{D}\gg h/\sqrt{2\pi m_{e}T}$ does not hold. This may imply that our derivation should be modified, since we explicitly assume that the reverse reaction is not affected by plasma screening, and $\langle\sigma v\rangle_{r}$ depends only on $T$. This assumption may no longer hold for $\lambda_{D}\sim h/\sqrt{2\pi m_{e}T}$ despite the fact that the reverse reaction involves an interaction with a neutral particle ($\nu_e$), since the Coulomb interaction of the outgoing particles ($pep$) and electron degeneracy should be taken into account in the calculation of the cross section. As a result, $\langle\sigma v\rangle_{r}$ may be affected. However, as explained below, detailed calculations for the plasma conditions near the centre of the Sun suggest that Equation~\eqref{eq:pep screening factor} holds to high accuracy (a few percents).

\citet{Gruzinov1997} calculated the rate of $^{7}$Be electron capture for solar conditions by numerically integrating the density matrix equation for a thermal electron in the field of a $^{7}$Be ion and other plasma ions and smeared out electrons. They found out that to within $1\%$ accuracy, the effect of screening can be described by the Salpeter enhancement. \citet{Simonucci2013} have later obtained similar results (to within a few percent agreement) using a different calculation technique. In fact, \citet{Gruzinov1997} calculated the deviation from the Salpeter enhancement for electron capture by ions with charges $Z\le6$, and specifically found $1\%$ accuracy for the $Z=2$ case, which is relevant for the second stage of the $pep$ reaction -- electron capture by $^2\mathrm{He}$. It follows from the discussion in Section~\ref{sec:proof}, that if the exact enhancement factor for the direct reaction equals the Salpeter enhancement, then the rate of the reverse reaction (neutrino capture) is unaffected by the screening. This suggests that $\langle\sigma v\rangle_{r}$ is density independent to a high (a few percent) accuracy for solar conditions.

For the conditions near the centre of the Sun, $T=1.6\times10^{7}\,\textrm{K}$, $\rho=150\,\textrm{g}\,\textrm{cm}^{-3}$, $X_{p}=0.36$, $X_{\textrm{He}}=0.64$, where $X_{i}$ are the mass fractions, we may take the limit of weak coupling and ignore the small degeneracy of the electrons. In this case the Coulomb interaction contribution to the chemical potential is given by the Debye--H\"{u}ckel theory,
\begin{equation}\label{eq:mu_c}
  \mu^C_p=\mu^C_e=-\frac{1}{2}\frac{e^2}{\lambda_D},
\end{equation}
where $\lambda_D$ is the Debye length,
\begin{equation}\label{eq:lD}
  \lambda_D^{-2}=\frac{4\pi e^2}{T}\left(n_e+\sum_a n_a Z_a^2\right).
\end{equation}
The summation includes all background plasma ions (the 'trace' interacting ions do not contribute). In this case both $\mu_p^{C}=\mu_e^{C}<0$, such that the $pep$ reaction is suppressed by $\exp(2\beta\mu_{p}^{C})$. This is the same amount by which the $pp$ reaction is enhanced, using Equation~\eqref{eq:screening factor}, $\myapprox4.8\%$. The $pep$ solar neutrino flux has been measured to $\sim20\%$ accuracy \citep{Bellini2012,Agostini2017}, and the required accuracy to measure the $pep$ screening suppression may be reached in the near future \citep{Newstead2018}.

Taking our estimate for the $pep$ screening at face value, the ratio between the screening factors of the $pp$ and the $pep$ reactions is larger than unity by $\myapprox10\,\%$, which may have consequences for a detailed modelling of the Sun. For example, \citet[][]{Bergstrom2016} used the ratio of the $pep$ neutrino flux to the $pp$ neutrino flux (with uncertainty of $\myapprox1\%$) as a constraint on their solar modelling, since these reactions have the same nuclear matrix element. However, their procedure assumes that these reactions also share the same screening, which is incorrect to the level of $\myapprox10\%$.

\section*{Acknowledgements}
We thank Andrei Gruzinov, Boaz Katz, Kfir Blum, Doron Gazit, Shimon Levit and Moti Milgrom for useful discussions. DK is supported by the Israel Atomic Energy Commission - The Council for Higher Education - Pazi Foundation and by a research grant from The Abramson Family Center for Young Scientists. EW is partially supported by Minerva, IMOS and ISF I-Core grants. AIC is partially supported by Fundamental Research program P12 of Presidium of Russian Academy of Sciences.






\bsp	
\label{lastpage}
\end{document}